\documentclass[12pt]{article}

\usepackage{amssymb}
\usepackage{amsmath}
\usepackage{latexsym}
\usepackage{epsfig}
\usepackage{graphicx}
\usepackage{subfigure}
\usepackage{wasysym}
\usepackage{threeparttable}
\usepackage{booktabs}
\usepackage[round]{natbib}
\usepackage{color}
\usepackage{epstopdf}
\usepackage{bm}
\usepackage{tabularx}
\usepackage{verbatim}
\usepackage{hyperref}
\usepackage{float}
\usepackage[ruled,vlined]{algorithm2e}
\usepackage{ulem}

\def\boxit#1{\vbox{\hrule\hbox{\vrule\kern6pt
          \vbox{\kern6pt#1\kern6pt}\kern6pt\vrule}\hrule}}

\def\bse{\begin{eqnarray*}}
\def\ese{\end{eqnarray*}}
\def\be{\begin{eqnarray}}
\def\ee{\end{eqnarray}}
\def\bq{\begin{equation}}
\def\eq{\end{equation}}
\def\bse{\begin{eqnarray*}}
\def\ese{\end{eqnarray*}}

\begin{document}
\thispagestyle{empty} \baselineskip=28pt

\begin{center}
{\LARGE{\bf Computationally Efficient Bayesian Unit-Level Models for Non-Gaussian Data Under Informative Sampling}}
%{\LARGE{\bf A Comprehensive Overview of Unit Level Modeling of Survey Data Under
%    Informative Sampling with Applications to Small Area Estimation}}

\end{center}

\baselineskip=12pt

\vskip 2mm
\begin{center}
Paul A. Parker\footnote{(\baselineskip=10pt to whom correspondence should be
  addressed) Department of Statistics, University of Missouri, 146 Middlebush Hall,
  Columbia, MO 65211-6100, paulparker@mail.missouri.edu}\,\footnote{Support
  for this research at the Missouri Research Data Center (MURDC) and through the Census Bureau Dissertation Fellowship program is gratefully acknowledged.}$^{5 7}$,
Scott H. Holan\footnote{\baselineskip=10pt Department of Statistics, University of
  Missouri, 146 Middlebush Hall, Columbia, MO 65211-6100,
  holans@missouri.edu}\,\footnote{\baselineskip=10pt Office of the Associate Director for Research and Methodology, U.S. Census Bureau, 4600 Silver
  Hill Road, Washington, D.C. 20233-9100, scott.holan@census.gov}\footnote{\baselineskip=10pt This research was partially supported by the
  U.S.~National Science Foundation (NSF) under NSF grant SES-1853096.}$^{7}$,\,
   and Ryan Janicki\footnote{\baselineskip=10pt Center for Statistical Research and
  Methodology, U.S. Census Bureau, 4600 Silver Hill Road, Washington,
  D.C. 20233-9100, ryan.janicki@census.gov}\footnote{\baselineskip=10pt   This article is released to inform interested parties of ongoing
  research and to encourage discussion. The views expressed on statistical issues are
  those of the authors and not those of the NSF or U.S. Census Bureau.
  The DRB approval number for this paper is CBDRB-FY20-355.}

\end{center}

\vskip 4mm

\begin{center}
  {\bf Abstract}
\end{center} Statistical estimates from survey samples have traditionally been obtained via design-based estimators. In many cases, these estimators tend to work well for quantities such as population totals or means, but can fall short as sample sizes become small. In today's “information age,” there is a strong demand for more granular estimates. To meet this demand, using  a Bayesian pseudo-likelihood, we propose a computationally efficient unit-level modeling approach for non-Gaussian data collected under informative sampling designs. Specifically, we focus on binary and multinomial data. Our approach is both multivariate and multiscale, incorporating spatial dependence at the area-level. We illustrate our approach through an empirical simulation study and through a motivating application to health insurance estimates using the American Community Survey.

\baselineskip=12pt 

\baselineskip=12pt
\par\vfill\noindent

\par\medskip\noindent

\clearpage\pagebreak\newpage \pagenumbering{arabic}
\baselineskip=12pt

\section{Introduction}\label{sec: intro}

An important dichotomy in the realm of small area estimation is that of area-level versus unit-level modeling approaches. In general, area-level models use the design-based direct estimate as a response within a statistical model. These models tend to smooth the noisy direct estimates in some fashion and estimate the true latent population value. In contrast to this, unit-level models treat the individual survey respondents as observations in the statistical model. Predictions can then be made for the entire population and aggregated as necessary to produce the desired estimates. As the need for more granular estimates becomes essential, area-level models may perform poorly due to underlying direct estimates with extremely small or nonexistent sample sizes. Unit-level approaches offer an attractive alternative by modeling the individual survey responses directly rather than smoothing the direct estimators. Although unit-level methodologies offer many advantages over their area-level counterparts, they also face their own set of challenges.

The primary difficulty with modeling survey data at the unit level is the consideration of informative sampling. Many surveys are sampled in an informative manner, whereby there is dependence between the probability of selection and the response of interest. When this relationship is not accounted for, increased bias may be present in the corresponding estimates \citep{pfe07}. The basic unit-level model, introduced by \citet{bat88}, assumes that the sample model holds for the entire population, and thus does not account for informative sampling.  \citet{par19} review the current methods for addressing the problem of informative sampling. Of primary interest is the pseudo-likelihood (PL) method \citep{ski89, bin83}, which exponentially weights each unit's likelihood contribution according to the corresponding survey weight. \citet{sav16} extend the PL approach to Bayesian settings and provide theoretical justification.

Although the problem of informative sampling has been studied in depth, there are other concerns with unit-level modeling that have received considerably less attention. In general, one major difference between area and unit-level approaches is dimensionality.  
Modeling survey data at the unit level can result in sample sizes that are magnitudes larger than those considered at the area level.  Unit-level models are fit to individual survey responses, which can number in the millions for large-scale surveys.  In contrast, area-level models are typically fit to aggregated survey statistics, such as survey-weighted means, which may number in the thousands.  For example, the American Community Survey (ACS) samples 3.5 million households annually, which may reasonably fall under the realm of ``big data." With these extremely large sample sizes comes computational concerns that must be addressed in order to make unit-level modeling viable. To further exacerbate the problem, many survey variables are non-Gaussian, which can lead to non-conjugate full conditional distributions when modeling dependence relationships using traditional Bayesian hierarchical models. Sampling from these posterior distributions can require Metropolis steps that are not efficient and can be cumbersome to tune.

\citet{brad20} introduce a class of conjugate prior distributions that may be used to model dependence for non-Gaussian data in the natural exponential family. This covers important cases such as Binomial, Multinomial, and Poisson data. \citet{parker20} extend this approach to model count data at the unit level under informative sampling, through the use of a PL. Unfortunately, sampling from the full conditional distributions can be difficult under these approaches when observations fall on the boundary of the data  (i.e. zero for Poisson data, zero or one for Bernoulli data, etc.). \citet{parker20} work around this by using an importance sampling scheme that works well when there are not an excessive number of boundary values (zeroes for Poisson data). However, many surveys contain a multitude of Binomial or Bernoulli random variables, which results in an abundance of boundary counts.

There are a number data augmentation approaches that have been developed to yield conjugate full-conditional distributions for Bernoulli data. \citet{albert1993bayesian} use latent Gaussian variables in conjunction with a probit link function to model Bernoulli data. More recently, \citet{pol13} use latent P\'{o}lya-Gamma random variables to model Binomial data with a logit link function. This approach may also be used to model Negative Binomial as well as Multinomial data.

In this paper, we develop methodology to model Binomial and Multinomial data at the unit level in a computationally efficient manner, while accounting for informative sampling. This is done through the use of Bayesian hierarchical modeling, in order to capture various sources of dependence. We consider both a Gibbs sampling approach with fully conjugate full conditional distributions, as well as a Variational Bayes approach to model fitting.

As a motivating example, we consider the problem of estimation of the proportion of people with health insurance at the county level for different income to poverty ratio (IPR) categories.  Currently, the Small Area Health Insurance Estimates (SAHIE) program within the U.S. Census Bureau produces estimates of health insurance rates using an area-level small area model fit to direct survey estimates using ACS data \citep{bau18}.  The model-based estimates produced by SAHIE are the only source of single year health insurance coverage estimates at the county level.  While the estimates are generally more precise than the corresponding direct estimates, there are serious modeling challenges with developing area level models for health insurance coverage.  First, there are boundary issues, in that many of the direct estimates at the county level are exactly equal to either 0 or 1, making use of continuous models impossible.  Second, there are policy requirements to benchmark lower-level county estimates to state-level estimates, so that users have confidence in the quality of the data.  Third, there are multiple within-county estimates that need to be produced, such as health insurance coverage by income level, and accounting for within-county dependencies in an area-level model can be difficult.  Finally, the computational requirements of fitting the model used by SAHIE are enormous, due to the complexity of the model and the number of estimates that are produced, despite the fact that an area-level model is used.

The model proposed in this paper eliminates many of these problems.  The boundary issues are resolved by using non-Gaussian likelihoods.  There is no need to benchmark estimates, as the PL produces predictions at the unit level, which can then be aggregated up to any desired geographic level.  Spatial and multivariate dependencies are handled throuh careful specification of the process model.  Finally, computational efficiency is achieved through a Variational Bayes approximation.

\begin{comment}
As a motivating example we consider estimates of health insurance rates by county, in a similar manner to the Small Area Health Insurance Estimates program (SAHIE). These estimates rely on data from the ACS, and thus models conducted at the unit level rely on extremely large samples, even though the sample size within each region of interest may be quite small.
\end{comment}

The remainder of this paper is organized as follows. Section \ref{sec: methods} introduces some necessary background material and then presents our proposed models as well as the methodology used to fit the models. We conduct an empirical simulation study in Section \ref{sec: sim}. We also provide a data analysis in Section \ref{sec: DA} where we estimate the health insurance rate for each county and five different income categories for the entire continental US. Finally, we provide concluding remarks and discussion in Section \ref{sec: disc}.

\section{Methodology}\label{sec: methods}

Models for small area estimation (SAE) often include area-level random effects in order to incorporate spatial dependence. These random effects are typically modeled using a latent Gaussian process (LGP), and Bayesian hierarchical modeling is a common technique used to fit these models. This may be computationally efficient when considering a Gaussian response, as it leads to conjugate full conditional distributions, however when the data model (likelihood) is non-Gaussian, sampling from the posterior distribution can become difficult as it may require the use of Metropolis type steps. These sampling mechanisms require tuning that can become unwieldy especially in high dimensional situations.

\citet{pol13} use a data augmentation scheme to allow for conjugate sampling under logistic likelihoods. Importantly, this includes both Bernoulli and Multinomial responses, which is useful as binary and categorical data are two often observed types of non-Gaussian survey data. This class also includes the Negative-Binomial distribution, which may be used to model count data.

Specifically, \citet{pol13} define a random variable $X$ to have a P\'{o}lya-Gamma distribution with parameters $b>0$ and $c \in \mathcal{R}$, denoted $\hbox{PG}(b,c)$, if $X$ is equal in distribution to
$$ \frac{1}{2\pi^2} \sum_{k=1}^{\infty} \frac{g_k}{(k-1/2)^2 + c^2/(4\pi^2)},$$ where $g_k \stackrel{ind}{\sim} Gamma(b,1)$. Furthermore, they show that 
\begin{equation}\label{eq: pg}
   \frac{(e^{\psi})^a}{(1 + e^{\psi})^b} = 2^{-b}e^{\kappa \psi} \int_0^{\infty} e^{-\omega \psi^2/2} p(\omega) d\omega, 
\end{equation} where $\kappa = a - b/2$ and $p(\omega)$ is a $\hbox{PG}(b,0)$ density. They also show that $(\omega | \psi) \sim \hbox{PG}(b,\psi)$. Thus, with a Binomial likelihood, using this data augmentation scheme and Gaussian prior distributions, one can sample from Gaussian full conditional distributions for the parameters, and P\'{o}lya-Gamma distributions for the latent augmentation variables. The \texttt{BayesLogit} package in \texttt{R} provides efficient sampling of P\'{o}lya-Gamma random variables \citep{windle2013bayeslogit}

\subsection{Pseudo-Likelihoods}

One of the main difficulties when implementing unit-level models for survey data is accounting for an informative sampling design. For example, certain demographic subgroups may be sampled with higher probability, but there may also be a relationship between these subgroups and the response variable of interest. Under this scenario, the sample is not representative of the population, and thus the sample likelihood should be adjusted to account for this. \citet{par19} give a review of modern methods for unit-level modeling under informative sampling. One general approach is to use a pseudo-likelihood, introduced by \cite{ski89} and \cite{bin83}, by weighting each unit's likelihood contribution using the reported survey weight $w_i$,

\begin{equation}\label{E: LogLikelihood}
  \prod_{i \in \mathcal{S}}  f( y_i \mid \boldsymbol{\theta})^{w_i},
\end{equation} where $\mathcal{S}$ indicates the sample and $y_i$ represents the response value for unit $i$.

The PL can be maximized using maximum-likelihood techniques, however \citet{sav16} show that a PL may also be used in a Bayesian setting, thus generating a pseudo-posterior distribution
$$\hat{\pi}(\bm{\theta} | \mathbf{y}, \mathbf{\tilde{w}}) \propto \left\{ \prod_{i \in \mathcal{S}} f(y_{i} | \bm{\theta})^{\tilde{w}_{i}} \right\} \pi (\bm{\theta}).$$ They emphasize the importance of scaling the weights to sum to the sample size, $\tilde{w}_i=n\frac{w_i}{\sum_{j=1}^n w_j}$, in order to prevent contraction of the PL and achieve appropriate variance estimates.

Using a unit-level model such as this, it is simple to generate predictions for any unobserved units, thereby effectively generating the population. It is then straightforward to aggregate units in order to estimate any finite population quantities, such as for SAE purposes. Under a Bayesian framework, this can be done for each sample from the posterior distribution, thus yielding a posterior distribution over any desired estimates. In the special case where all covariates are categorical in nature, this approach can be seen as a type of poststratification \citep{gelman97, park06}. \citet{zha14} provide an example of a pseudo-likelihood and poststratification combination for small area estimates in a frequentist framework, whereas \citet{parker20} take a Bayesian pseudo-likelihood and poststratification approach.

Now, an unweighted binomial likelihood has the form
$$
\prod_{i \in \mathcal{S}} \frac{(e^{\psi_i})^{y_i}}{(1 + e^{\psi_i})^{n_i}}.
$$ By using a pseudo-likelihood instead, the form becomes
\begin{equation}\label{eq: PL_bin}
   \prod_{i \in \mathcal{S}} \left(\frac{(e^{\psi_i})^{y_i}}{(1 + e^{\psi_i})^{n_i}}\right)^{\tilde{w}_i} =
           \prod_{i \in \mathcal{S}} \frac{(e^{\psi_i})^{y_i^*}}{(1 + e^{\psi_i})^{n_i^*}},
\end{equation} where $y_i^*=y_i \times \tilde{w}_i$ and $n_i^*=n_i \times \tilde{w}_i$. The PL given by (\ref{eq: PL_bin}) is of the same form as that given in (\ref{eq: pg}), thus we are able to sample from conjugate full conditional distributions using a binomial type PL with Guassian prior distributions, and PG data augmentation variables.

\subsection{Binomial Response Model}

Using the P\'{o}lya-Gamma data augmentation scheme, we develop a computationally efficient pseudo-likelihood mixed model for binomial survey data (PL-MB) under informative sampling,
    \begin{equation}
    \begin{split}
        \bm{Z} | \bm{\beta, \eta} & \propto \prod_{i \in S} \hbox{Bin}\left(Z_i | n_i, p_i \right)^{\stackrel{\sim}{w}_i} \\
        \hbox{logit}(p_i) &= \bm{x_i'} \bm{\beta} + \bm{\phi_i'} \bm{\eta} \\
        \bm{\eta}|\sigma^2_{\eta} & \sim \hbox{N}_r(\bm{0_r}, \sigma_{\eta}^2 \bm{I}_r ) \\
        \bm{\beta} & \sim  \hbox{N}_q(\bm{0_q}, \sigma_{\beta}^2 \bm{I}_q ) \\
        \sigma_{\eta}^2 & \sim \hbox{IG}(a, b) \\
        & \sigma_{\beta},  a, b >0,
    \end{split} 
\end{equation} where $Z_i$ represents the response for unit $i \in \mathcal{S}$. We model the data using a Binomial pseudo-likelihood, with $n_i$ representing the number of trials, and $p_i$ representing the probability of success under each trial. In many survey data scenarios, including those explored here, the data is binary, thus $n_i=1, \forall i$.  The vector $\bm{x_i'}$ represents a $q$-dimensional set of covariates and $\bm{\beta}$ is the $q$-dimensional vector of fixed effects. In this work, the vector $\bm{\phi_i'}$ represents either an $r$-dimensional vector of spatial basis functions, or an incidence vector, indicating which area unit $i$ resides in. In this way, the $r$-dimensional vector $\bm{\eta}$ act as area-level random effects. The full conditional distributions for Gibbs sampling, which rely on a P\'{o}lya-Gamma data augmentation scheme, can be found in the Appendix.

\subsection{Variational Bayes Approximation}

In many high-dimensional settings, it can become a computational burden to sample from the posterior distribution via MCMC, even through the use of Gibbs sampling with fully conjugate full conditional distributions. For example, using the P\'{o}lya-Gamma data augmentation scheme, a latent random variable must be drawn for every sample observation at every iteration of the MCMC. As sample sizes become very large, this may become infeasible, even after allowing for parallel computing techniques. One popular solution to this computational problem is the variational Bayes approach \citep{jordan99, wainwright08}, for which an approximation to the posterior distribution is used rather than the true posterior distribution. A class of distributions, $\mathcal{D}$, is chosen for $q^*(\bm{\theta})$, the approximation to the true posterior, $p(\bm{\theta} | \bm{x})$. Optimization techniques may then be used to minimize the Kullback-Leibler (KL) divergence between the approximate and true posterior distributions,
\begin{equation}
    q^*(\bm{\theta}) = \hbox{arg\,min}_{q(\bm{\theta}) \in \mathcal{D}} \hbox{KL}\left(q(\bm{\theta}) || p(\bm{\theta} | \bm{x})\right).
\end{equation}

\citet{beal03} focus on a specific case known as the variational Bayes EM algorithm. The approximating distribution can be factored into a product of global parameters and local latent variables, $q(\bm{\theta})=q(\bm{\beta})\prod_{i=1}^n q(\xi_i)$. With this factorization, an iterative approach can be used to minimize the KL divergence, where 
\begin{equation}
    \begin{split}
        q(\bm{\beta})^{(t)} &\propto \hbox{exp}\left\{\mathbb{E}_{q^{(t-1)}(\bm{\xi}))} \hbox{log}[p(\bm{\beta}|\bm{Z},\bm{\xi})] \right\} \\ 
        q(\xi_i)^{(t)} &\propto \hbox{exp}\left\{\mathbb{E}_{q^{(t-1)}(\bm{\beta}))} \hbox{log}[p(\xi_i|\bm{Z}, \bm{\xi}_{-i},\bm{\beta})] \right\}, \; i=1,\ldots,n.
    \end{split}
\end{equation} In models that use fully conjugate full conditional distributions, as well as likelihoods from the exponential family, these factorized approximate distributions are of the same class as their corresponding full conditional distribution. Importantly, this includes the case of logistic regression via P\'{o}lya-Gamma data augmentation, for which \citet{durante19} explore a variational Bayes EM algorithm approach.

Algorithm \ref{A: one} provides an extension of the one explored by \citet{durante19}. The main extension of this algorithm is the inclusion of the pseudo-likelihood rather than the original Binomial likelihood. This algorithm may be used in place of MCMC in order to fit the PL-MB model in high dimensional settings. Independent samples from the variational approximation to the posterior of $\bm{\zeta}=(\bm{\beta'}, \bm{\eta'})$ may be drawn by sampling from a $\hbox{N}(\tilde{\bm{\mu}}, \tilde{\bm{\Sigma}})$ distribution, which may then be used to produce any desired Monte Carlo estimates.
\\

\begin{algorithm}[H]
\label{A: one}
\SetAlgoLined

 Initialize $\tilde{\sigma}^2_{\eta}$ and $\tilde{\xi}_i, \; i=1,\ldots,n$ \;
 Let $\bm{D}=\left[\bm{X}, \bm{\Phi} \right]$ and $\bm{\zeta}=(\bm{\beta'}, \bm{\eta'})$ \;
 \For{$t=1$ until convergence}{
  $\tilde{\bm{\Omega}}=\hbox{Diag}\left(\frac{w_1}{2\tilde{\xi}_1}\hbox{tanh}(\tilde{\xi}_1/2),\ldots,\frac{w_n}{2\tilde{\xi}_n}\hbox{tanh}(\tilde{\xi}_n/2)\right)$\;
  $\tilde{\bm{\Sigma}} =\left(\hbox{blockdiag}(\frac{1}{\sigma^2_{\beta}} \bm{I}_p, \frac{a+r/2}{\tilde{\sigma}^2_\eta} \bm{I}_r) + \bm{D'}\tilde{\bm{\Omega}}\bm{D} \right)^{-1}$\;
  $\tilde{\bm{\Sigma}}_{\eta} = \tilde{\bm{\Sigma}}[(p+1):(p+r), (p+1):(p+r)]$\;
  $\tilde{\bm{\mu}}=(\tilde{\bm{\mu}}_{\beta}', \tilde{\bm{\mu}}_{\eta}')'=\tilde{\bm{\Sigma}}\bm{D'}\left(\bm{w}\odot(\bm{Z} - 1/2)  \right) $\;
  $\tilde{\sigma}^2_{\eta}=b + \frac{1}{2}\left(\tilde{\bm{\mu}}_{\eta}'\tilde{\bm{\mu}}_{\eta} + \hbox{tr}(\tilde{\bm{\Sigma}}_{\eta}) \right)$\;
  \For{$i=1$ to $n$}{
  $\tilde{\xi}_i = \left(\bm{D}_i'\tilde{\bm{\Sigma}}\bm{D}_i + (\bm{D}_i'\tilde{\bm{\mu}})^2\right)^{1/2}$\;
  }
  
 }
 \caption{VB EM algorithm for PL-MB model}
\end{algorithm}

\subsection{Multinomial Response Model}

In addition to Binomial data, Multinomial or categorical data is often observed in survey data. In a similar fashion as the PL-MB model, we can write the Pseudo-likelihood mixed effect Multinomial model (PL-MM) with $K$ categories as 
\begin{equation}
    \begin{split}
                \bm{Z} | \bm{\beta, \eta} & \propto \prod_{i \in S} \hbox{Multinomial}\left(\bm{Z_i} | n_i, \bm{p_i} \right)^{\stackrel{\sim}{w}_i} \\
                p_{ik} &= \frac{\hbox{exp}(\psi_{ik})}{\sum_{k=1}^K \hbox{exp}(\psi_{ik})} \\
        \psi_{ik} &= \bm{x_i'} \bm{\beta_k} + \bm{\phi_i'} \bm{\eta_k} \\
        \bm{\eta_k}|\sigma^2_{\eta k} & \sim \hbox{N}_r(\bm{0_r}, \sigma_{\eta k}^2 \bm{I}_r ), \; k=1,\ldots,K-1 \\
        \bm{\beta_k} & \sim  \hbox{N}_p(\bm{0_p}, \sigma_{\beta}^2 \bm{I}_p ), \; k=1,\ldots,K-1 \\
        \sigma_{\eta k}^2 & \sim \hbox{IG}(a, b), \; k=1,\ldots,K-1 \\ 
        & \sigma_{\beta},  a, b >0,
    \end{split}
\end{equation} where $\bm{\beta_K}$ and $\bm{\eta_K}$ are constrained to be equal to zero for identifiability. The $K$-dimensional vector $\bm{Z_i}$ represents the number of successful outcomes in each of the $K$ categories for survey unit $i$, and the $K$-dimensional vector $\bm{p}_i$ represents the probability of each category for unit $i$.

Although Algorithm \ref{A: one} is intended for Binomial data, a stick-breaking representation of the Multinomial distribution can be used to expand the applicability of this VB approach. Specifically, \citet{linderman15} show that the Multinomial distribution may be written as a product of independent Binomial distributions,
\begin{equation}
    \begin{split}
        \hbox{Multinomial}(\bm{Z}|n, \bm{p})=\prod_{k=1}^{K-1} \hbox{Bin}(Z_k|n_k, \tilde{p}_k),
    \end{split}
\end{equation} where
\begin{equation}
    n_k = n - \sum_{j < k}Z_j, \; \; \tilde{p}_k = \frac{p_k}{1 - \sum_{j<k} p_j}, \; \; k=2,\ldots,K.
\end{equation} Under this view of Multinomial data, we can rewrite the PL-MM model as 
\begin{equation}
    \begin{split}
                \bm{Z} | \bm{\beta, \eta} & \propto \prod_{i \in S} \prod_{k=1}^{K-1} \hbox{Bin}\left(Z_{ik} | n_{ik}, \tilde{p}_{ik} \right)^{\stackrel{\sim}{w}_i} \\
        \hbox{logit}(\tilde{p}_{ik}) &= \bm{x_i'} \bm{\beta_k} + \bm{\phi_i'} \bm{\eta_k} \\
        \bm{\eta_k}|\sigma^2_{\eta k} & \sim \hbox{N}_r(\bm{0_r}, \sigma_{\eta k}^2 \bm{I}_r ), \; k=1,\ldots,K-1 \\
        \bm{\beta_k} & \sim  \hbox{N}_p(\bm{0_p}, \sigma_{\beta}^2 \bm{I}_p ), \; k=1,\ldots,K-1 \\
        \sigma_{\eta k}^2 & \sim \hbox{IG}(a, b), \; k=1,\ldots,K-1 \\
        & \sigma_{\beta},  a, b >0,
    \end{split}
\end{equation} where $n_{ik}=n_i - \sum_{j < k}Z_{ij}$ and $\tilde{p}_{ik} = \frac{p_{ik}}{1 - \sum_{j<k} p_{ij}}, \; \; k=2,\ldots,K$. Thus, the PL-MM model may be fit as a series of $K-1$ independent Binomial models using either MCMC or the VB approach outlined in Algorithm \ref{A: one}. Note that after fitting the model, the stick breaking probabilities $\bm{\tilde{p}}_i$ can be transformed back to the original probabilities $\bm{p}_i$ for inference.

\section{Empirical Simulation Study}\label{sec: sim}

In order to mimic a real survey data setting, our simulations revolve around resampling of an existing survey dataset rather than generating a synthetic population from a parametric distribution. Specifically, we treat the existing survey sample as our population and then take a further sample with probability proportional to $s_i$, a size variable that is constructed in an informative manner. This informative sampling scheme can be validated by comparing the weighted design-based estimator to an unweighted design-based estimator. Under an informative design, the unweighted estimator will result in greater bias.

\subsection{Multinomial Response Simulation}

An important SAE application is the Small Area Health Insurance Estimation (SAHIE) program \citep{bau18}. The goal of SAHIE is to estimate the proportion of individuals with health insurance by county for a number of income to poverty ratio (IPR) categories. The number of people within each IPR category is unknown, and thus to create these estimates, health insurance and IPR category must be modeled simultaneously. The IPR categories under consideration are (0-138\%, 138-200\%, 200-250\%, 250-400\%, 400+\%), and within each IPR category, an individual may be categorized as either having or not having health insurance.  The choice of these IPR categories is motivated by the needs of one of SAHIE's sponsors, the Centers for Disease Control and Prevention (CDC), which provides cancer screenings for low income, uninsured women \citep{bau18}.
In this manner, we view individuals as falling into one of 10 distinct categories, $(C_{1,0},\ldots,C_{5,0},C_{1,1},\ldots,C_{5,1})$, where $C_{j,k}$ indicates an individual in IPR category $j=1,\ldots,5$ and health insurance indicator $k=0,1$.

To construct health insurance estimates by county and IPR category, we fit the PL-MM with ten categories using $n_i=1$ for all $i$. We let $\boldsymbol{x}_i$ consist of poststratification variables including race category, sex, and age category. We also let $\phi_i$ be a vector indicating which county unit $i$ resides in. Thus, the model uses a county level random effect. We use the vague prior distribution $\sigma^2_{\beta}=1000$ and $a=b=0.5$. The model is fit using both the MCMC and VB fitting strategies, with both drawing a posterior sample size of 1000, after discarding 1000 draws as burnin for MCMC. For MCMC, convergence was assessed visually through the use of traceplots of the sample chains, for which no lack of convergence was detected. After fitting the model on the sample data, predictions are made for all units in the population. The synthesized population is then aggregated to the desired level of the estimates (i.e., county by IPR category). This is done for each posterior draw, giving a posterior predictive distribution for the desired estimates.

To assess the SAE capability of our PL-MM model through simulation, we treat the 2014 1-year American Community Survey (ACS) sample in Minnesota as our population. This data contains roughly 120,000 respondents across Minnesota's 87 counties. We then take a further probability proportional to size sample without replacement, using the Poisson method \citep{brewer1984poisson} with an expected sample size of 10,000. We use the size variable $s_i=\hbox{exp}\left\{w^*_i+2 \hbox{I}(H_i=0)\right\}$, where $w^*_i$ is the original survey weight for unit $i$ after scaling to have mean zero and standard deviation of one, and $H_i$ indicates whether or not unit $i$ had health insurance. Estimates are constructed using the PL-MM with both MCMC and VB fits. We also construct a Horvitz-Thompson direct estimate as well as an unweighted direct estimate. We repeat the sampling and estimation process 50 times in order to compare MSE and bias across estimators.

A summary of the simulation results in given in Table \ref{tab: sim}, including average mean squared error (MSE) and squared bias for the competing estimators, as well as computation time and 95\% credible interval (CI) coverage rates for the two model based estimators. The higher bias of the UW estimator relative to the direct estimator indicates that the sampling scheme was indeed informative. The two model based approaches yield significant reductions to MSE when compared to the direct estimator. Surprisingly, the model fit with VB resulted in even lower MSE than the model fit using MCMC. This is likely because the non-Gaussian posterior of the MCMC approach allows for heavier tails which can influence the posterior predictive mean of the estimates. The downside to the VB approach is that the approximate posterior results in uncertainty estimates that are not optimal. This is reflected in the lower 95\% CI coverage rate for the VB approach compared to the MCMC approach. This is to be expected, as the VB approach only approximates the true posterior distribution. However, the differences are relatively minor, and can be justified through the massive decrease in computation time.

\begin{table}[H]
\begin{center}
 \begin{tabular}{||c | c c c c||} 
 \hline
 Estimator & MSE & Bias$^2$ & Time (s) & Coverage Rate  \\ [0.5ex] 
 \hline\hline
 Model MCMC & $7.1 \times 10^{-3}$ & $3.7 \times 10^{-3}$ & 7314 & $\bm{94\%}$ \\ 
 \hline
 Model VB & $\bm{2.3 \times 10^{-3}}$ & $\bm{1.7 \times 10^{-3}}$ & $\bm{140}$ & 87\% \\
 \hline
 Direct & $9.9 \times 10^{-2}$ & $3.8 \times 10^{-2}$ & - & -  \\
 \hline
 UW Direct & $1.6 \times 10^{-1}$ & $1.1 \times 10^{-1}$ & - & -  \\ [1ex] 
 \hline
\end{tabular}
\caption{MSE and squared bias of the four estimators averaged across counties based on simulation results. Average computation time in seconds and 95\% credible interval coverage rate are also given for the model based estimates.}
\label{tab: sim}
\end{center}
\end{table}

\begin{figure}[H]
    \begin{center}
        \includegraphics[width=150mm]{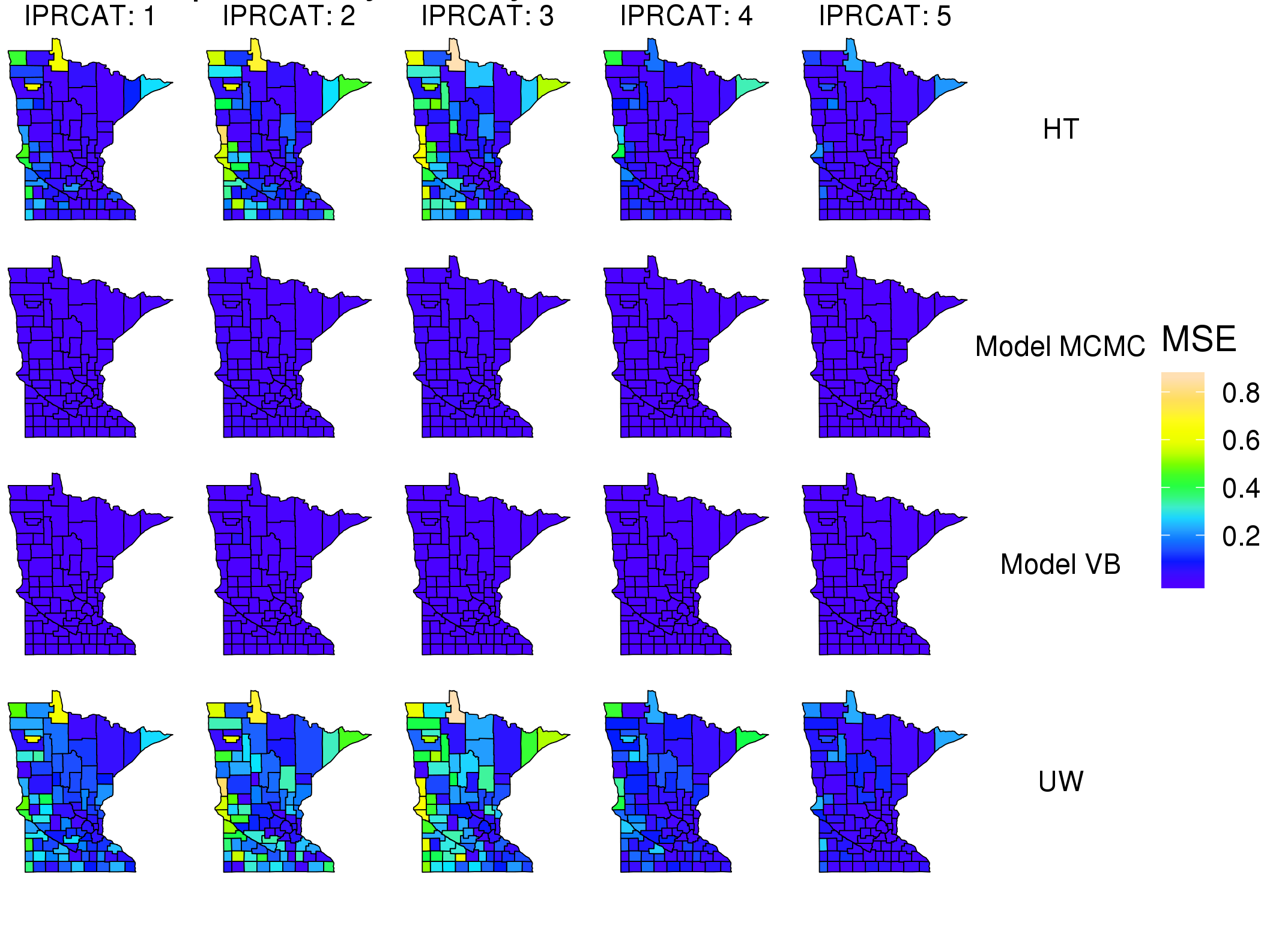}
         \caption{\baselineskip=10pt Emprical mean squared error by county across the simulation based estimates for the state of Minnesota. Columns represent the different IPR categories and rows represent the different estimators.}
         \label{fig: sim_mse}
    \end{center}               
\end{figure}

We also show the MSE by county and IPR category for each estimator in Figure \ref{fig: sim_mse}. The largest reductions in MSE through model-based estimation tend to occur for the more rural and sparsely populated regions of the state. These counties tend to have smaller sample sizes resulting in more erratic direct estimates. The model-based estimates borrow strength from sampled units in all counties, resulting in more stable estimates. 

\section{Data Analysis}\label{sec: DA}
The simulation in Section \ref{sec: sim} illustrates how the PL-MM model may be used to generate SAHIE type estimates for a single state. However, the SAHIE program is tasked with creating estimates for the entirety of the US rather than a single state. The bottleneck in the MCMC approach to the PL-MM model is the generation of P\'{o}lya-Gamma random variables for every sample observation at every MCMC iteration. Although this approach is feasible at a state level, it becomes unwieldy at the national level, where the ACS samples 3.5 million households annually. For this reason, we rely on the VB approach to the PL-MM model in order to create estimates of health insurance by county and IPR category for the entire continental US.

Again, we use the PL-MM model with 10 categories and $n_i=1$ for all $i$. We also use the same prior distribution and poststratification variables that were considered in Section \ref{sec: sim}. There are over 3,000 counties in the US, compared to only 87 in Minnesota, thus, we require a form of dimension reduction for $\phi_i$ rather than using county indicators. To do this, we let $\phi_i$ be equal to a set of spatial basis functions evaluated for unit $i$. Specifically, we use the first 307 (10\%) eigenvectors of the county adjacency matrix as our spatial basis functions. This choice was motivated in part by the suggestion of \citet{hughes2013dimension} to use 10\% of the available eigenvectors, as well as by the need for substantial dimension reduction with respect to the random effects.

\begin{figure}[H]
    \begin{center}
        \includegraphics[width=150mm]{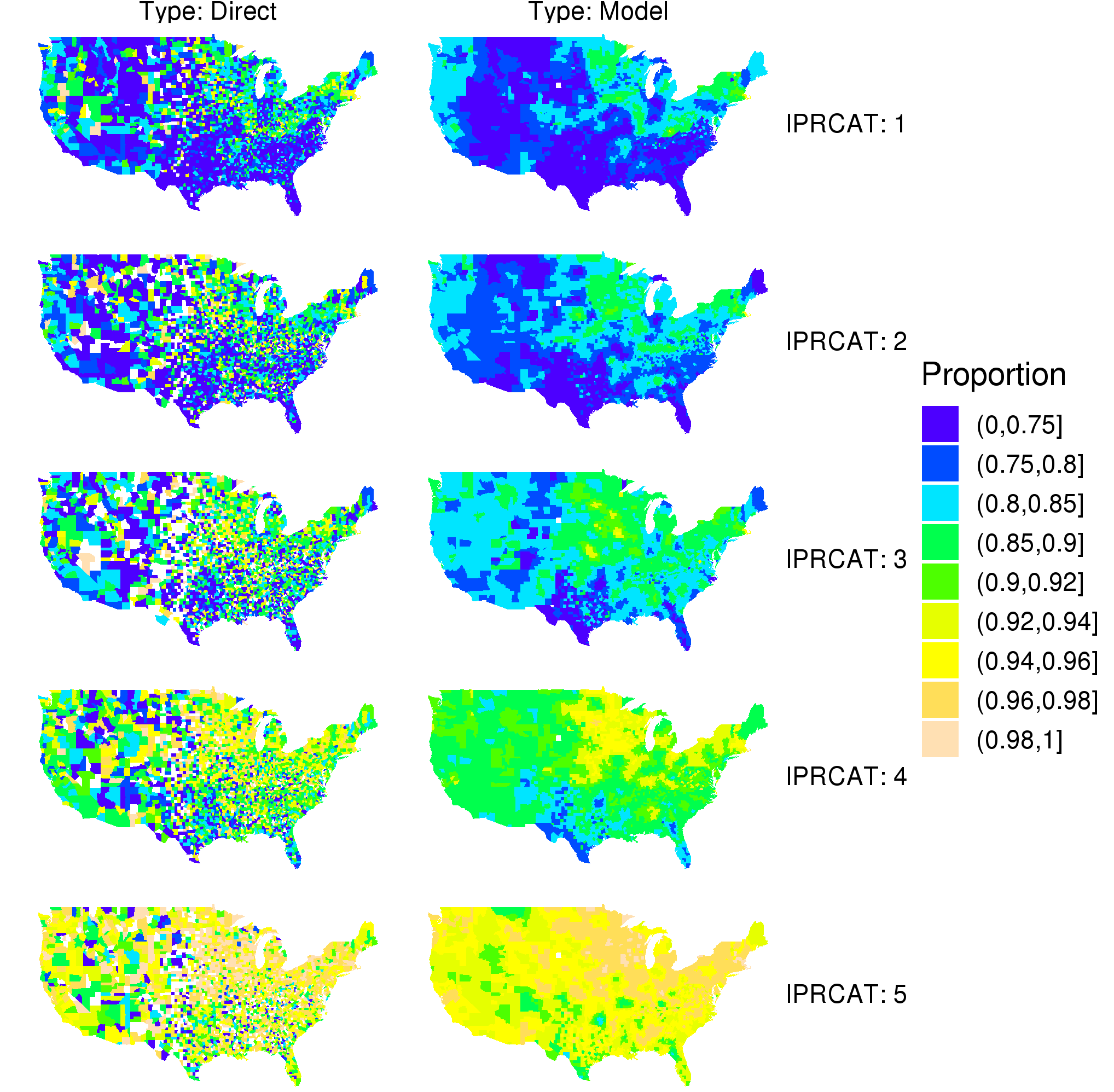}
         \caption{\baselineskip=10pt Direct and model based estimates of the proportion of the population with health insurance by county and IPR category for the continental United States.}
         \label{fig: FC_est}
    \end{center}               
\end{figure}

\begin{figure}[H]
    \begin{center}
        \includegraphics[width=150mm]{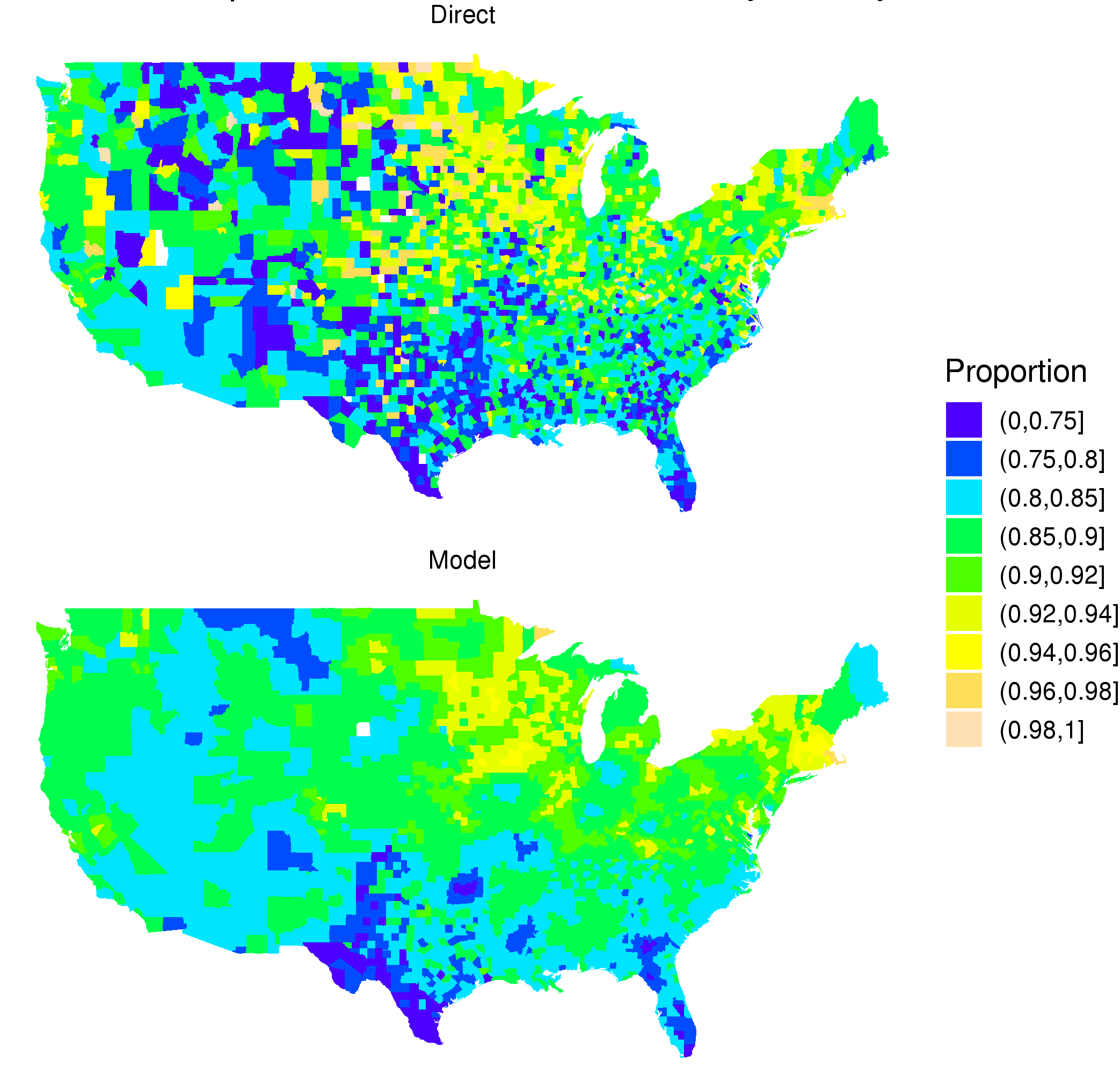}
         \caption{\baselineskip=10pt Direct and model based estimates of the proportion of the population with health insurance by county for the continental United States.}
         \label{fig: FC_est_cty}
    \end{center}               
\end{figure}

We fit the PL-MM model using the VB approach, with a sample size of roughly 4.5 million. We then take 1000 independent draws from the variational posterior distribution in order to construct the posterior predictive distribution of our estimates. Treating the posterior predictive mean as our point estimates, we plot the model based estimates alongside the direct estimates in Figure \ref{fig: FC_est}. Note that the direct estimates shown here have been infused with a small amount of noise in order to preserve respondent confidentiality. Visually, the direct estimates are quite noisy, due to the very small sample sizes in many counties. The model based estimates are able to provide a degree of smoothing through the use of borrowed information in the hierarchical model structure. This results in model based estimates that have the same general spatial pattern as the direct estimates without as much noise. We also plot the health insurance estimates by county without regard to IPR category in Figure \ref{fig: FC_est_cty}. Similar patterns can be noticed here.

\begin{figure}[H]
    \begin{center}
        \includegraphics[width=150mm]{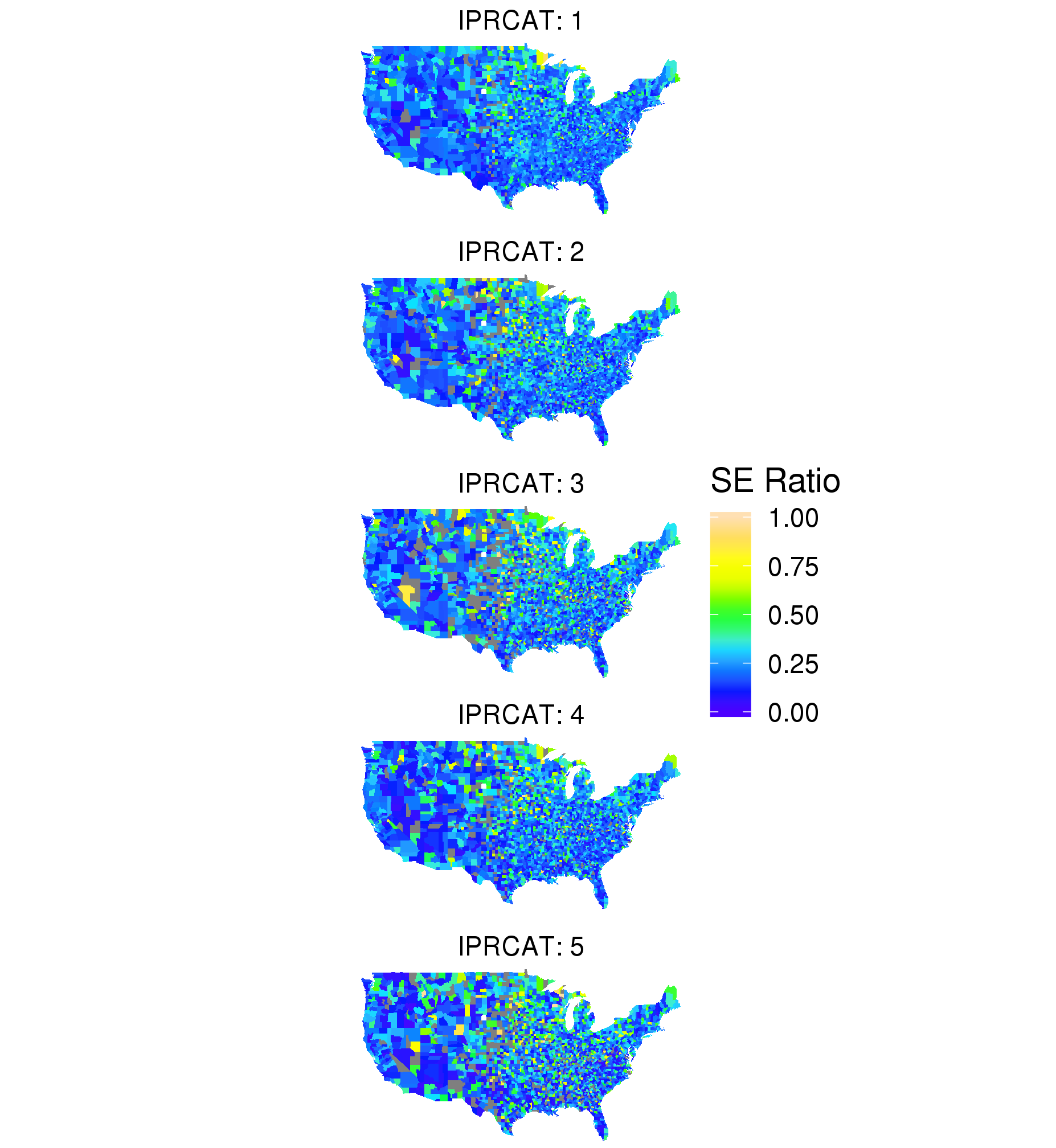}
         \caption{\baselineskip=10pt Ratio of model based standard errors to direct estimate standard errors by county and IPR category for the continental United States. Counties with no available direct estimate are shown in gray.}
         \label{fig: se}
    \end{center}               
\end{figure}

We plot the ratio of the model based standard errors to the direct estimate standard errors by county and IPR category in Figure \ref{fig: se}. For the vast majority of estimates, the model based approach provides quite substantial reductions in standard error, with the largest advantage occurring in the more sparsely populated Southern and Western regions of the country.

This example demonstrates how the PL-MB and PL-MM models may be used to model complex dependence structures with non-Gaussian data in a computationally efficient manner. The VB approach specifically was able to generate estimates for over 15,000 county and IPR category combinations, utilizing a sample size of over 4 million, in roughly 17 hours. These estimates are much less noisy than direct estimates, with substantially lower standard errors. Furthermore, the simulation results of Section \ref{sec: sim} indicate that these model based estimates should have much lower MSE. In addition to advantages over the direct estimate, this approach has many advantages over area-level modeling approaches, such as the one currently in use for SAHIE. For example, unit-level models allow for easy aggregation to multiple domains. A single PL-MM model may be used to give county and state level estimates, whereas area-level modeling strategies require two separate models and often rely on ad-hoc benchmarking techniques. Another advantage is that unit-level models do not require a direct estimate for a given area in order to construct an estimate, in contrast to area-level models.

\section{Discussion}\label{sec: disc}

This paper establishes a framework for modeling Binomial and Multinomial unit-level survey data, specifically under an informative sample. We envision this methodology being used to create area-level estimates of population proportions, with health insurance (SAHIE) as our motivating example. The current methodology used to generate SAHIE estimates is conducted at the area level which can cause a number of problems that are alleviated through the use of unit-level modeling. Our unit-level approach is able to generate multiple levels of estimates through a single model without the need for benchmarking techniques. We demonstrate this by producing health insurance estimates by county as well as by IPR category within each county for the entire continental US. Our approach is also able to produce very precise estimates compared to traditional direct estimators, as demonstrated by our empirical simulation study. Finally, these estimates can be produced in a very computationally efficient manner either through the use of either Gibbs sampling with fully conjugate full-conditional distributions or through a VB approximation to the posterior distribution.

Although this paper provides a methodological step forward for small area estimates of health insurance, further work would be necessary to create estimates that might replace the current SAHIE program. For example, the current SAHIE methodology considers a number of important covariates that were not considered here, due to disclosure limitations, including data from the Supplemental Nutrition Assistance Program as well as Medicaid. Furthermore, the method considered here is a type of generalized linear model, but there is potential for improvement through the use nonlinear modeling techniques, which is the subject of future work.

\clearpage\pagebreak\newpage

\baselineskip=14pt %\vskip 2mm\noindent
\bibliographystyle{jasa}
\bibliography{pg}

\begin{thebibliography}{22}
\newcommand{\enquote}[1]{``#1''}
\expandafter\ifx\csname natexlab\endcsname\relax\def\natexlab#1{#1}\fi

\bibitem[\protect\citename{Albert and Chib, }1993]{albert1993bayesian}
Albert, J.~H. and Chib, S. (1993).
\newblock \enquote{Bayesian analysis of binary and polychotomous response
  data.}
\newblock {\em Journal of the American statistical Association\/}, 88, 422,
  669--679.

\bibitem[\protect\citename{Battese et~al., }1988]{bat88}
Battese, G.~E., Harter, R.~M., and Fuller, W.~A. (1988).
\newblock \enquote{An error-components model for prediction of county crop
  areas using survey and satellite data.}
\newblock {\em Journal of the American Statistical Association\/}, 83, 401,
  28--36.

\bibitem[\protect\citename{Bauder et~al., }2018]{bau18}
Bauder, M., Luery, D., and Szelepka, S. (2018).
\newblock \enquote{Small area estimation of health insurance coverage in 2010
  -- 2016.}
\newblock Tech. rep., Small Area Methods Branch, Social, Economic, and Housing
  Statistics Division, U. S. Census Bureau.

\bibitem[\protect\citename{Beal and Ghahramani, }2003]{beal03}
Beal, M.~J. and Ghahramani, Z. (2003).
\newblock \enquote{The variational Bayesian EM algorithm for incomplete data:
  with application to scoring graphical model structures.}
\newblock {\em Bayesian statistics\/}, 7, 453--464.

\bibitem[\protect\citename{Binder, }1983]{bin83}
Binder, D.~A. (1983).
\newblock \enquote{On the variances of asymptotically normal estimators from
  complex surveys.}
\newblock {\em International Statistical Review\/}, 51, 3, 279--292.

\bibitem[\protect\citename{Bradley et~al., }2020]{brad20}
Bradley, J.~R., Holan, S.~H., and Wikle, C.~K. (2020).
\newblock \enquote{Bayesian Hierarchical Models with Conjugate Full-Conditional
  Distributions for Dependent Data from the Natural Exponential Family.}
\newblock {\em Journal of the American Statistical Association\/},  To Appear.

\bibitem[\protect\citename{Brewer et~al., }1984]{brewer1984poisson}
Brewer, K., Early, L., and Hanif, M. (1984).
\newblock \enquote{Poisson, modified Poisson and collocated sampling.}
\newblock {\em Journal of Statistical Planning and Inference\/}, 10, 1, 15--30.

\bibitem[\protect\citename{Durante et~al., }2019]{durante19}
Durante, D., Rigon, T., et~al. (2019).
\newblock \enquote{Conditionally conjugate mean-field variational Bayes for
  logistic models.}
\newblock {\em Statistical Science\/}, 34, 3, 472--485.

\bibitem[\protect\citename{Gelman and Little, }1997]{gelman97}
Gelman, A. and Little, T.~C. (1997).
\newblock \enquote{Poststratification into many categories using hierarchical
  logistic regression.}
\newblock {\em Survey Methodology\/}, 23, 127--135.

\bibitem[\protect\citename{Hughes and Haran, }2013]{hughes2013dimension}
Hughes, J. and Haran, M. (2013).
\newblock \enquote{Dimension reduction and alleviation of confounding for
  spatial generalized linear mixed models.}
\newblock {\em Journal of the Royal Statistical Society: Series B (Statistical
  Methodology)\/}, 75, 1, 139--159.

\bibitem[\protect\citename{Jordan et~al., }1999]{jordan99}
Jordan, M.~I., Ghahramani, Z., Jaakkola, T.~S., and Saul, L.~K. (1999).
\newblock \enquote{An introduction to variational methods for graphical
  models.}
\newblock {\em Machine learning\/}, 37, 2, 183--233.

\bibitem[\protect\citename{Linderman et~al., }2015]{linderman15}
Linderman, S., Johnson, M.~J., and Adams, R.~P. (2015).
\newblock \enquote{Dependent multinomial models made easy: Stick-breaking with
  the P{\'o}lya-Gamma augmentation.}
\newblock In {\em Advances in Neural Information Processing Systems\/},
  3456--3464.

\bibitem[\protect\citename{Park et~al., }2006]{park06}
Park, D.~K., Gelman, A., and Bafumi, J. (2006).
\newblock \enquote{State-level opinions from national surveys:
  Poststratification using multilevel logistic regression.}
\newblock {\em Public Opinion in State Politics\/}.

\bibitem[\protect\citename{Parker et~al., }2020]{parker20}
Parker, P.~A., Holan, S.~H., and Janicki, R. (2020).
\newblock \enquote{Conjugate Bayesian Unit-level Modeling of Count Data Under
  Informative Sampling Designs.}
\newblock {\em Stat\/},  e267.

\bibitem[\protect\citename{Parker et~al., }2019]{par19}
Parker, P.~A., Janicki, R., and Holan, S.~H. (2019).
\newblock \enquote{Unit level modeling of survey data for small area estimation
  under informative sampling: A comprehensive overview with extensions.}
\newblock {\em arXiv preprint arXiv:1908.10488\/}.

\bibitem[\protect\citename{Pfeffermann and Sverchkov, }2007]{pfe07}
Pfeffermann, D. and Sverchkov, M. (2007).
\newblock \enquote{Small-area estimation under informative probability sampling
  of areas and within the selected areas.}
\newblock {\em Journal of the American Statistical Association\/}, 102, 480,
  1427--1439.

\bibitem[\protect\citename{Polson et~al., }2013]{pol13}
Polson, N.~G., Scott, J.~G., and Windle, J. (2013).
\newblock \enquote{Bayesian inference for logistic models using
  P{\'o}lya--Gamma latent variables.}
\newblock {\em Journal of the American statistical Association\/}, 108, 504,
  1339--1349.

\bibitem[\protect\citename{Savitsky and Toth, }2016]{sav16}
Savitsky, T.~D. and Toth, D. (2016).
\newblock \enquote{Bayesian estimation under informative sampling.}
\newblock {\em Electronic Journal of Statistics\/}, 10, 1, 1677--1708.

\bibitem[\protect\citename{Skinner, }1989]{ski89}
Skinner, C.~J. (1989).
\newblock \enquote{Domain means, regression and multivariate analysis.}
\newblock In {\em Analysis of Complex Surveys\/}, eds. C.~J. Skinner, D.~Holt,
  and T.~M.~F. Smith,  80 -- 84. Chichester: Wiley.

\bibitem[\protect\citename{Wainwright et~al., }2008]{wainwright08}
Wainwright, M.~J., Jordan, M.~I., et~al. (2008).
\newblock \enquote{Graphical models, exponential families, and variational
  inference.}
\newblock {\em Foundations and Trends{\textregistered} in Machine Learning\/},
  1, 1--2, 1--305.

\bibitem[\protect\citename{Windle et~al., }2013]{windle2013bayeslogit}
Windle, J., Polson, N., and Scott, J. (2013).
\newblock \enquote{BayesLogit: Bayesian logistic regression.}
\newblock {\em URL http://cran. r-project. org/web/packages/BayesLogit/index.
  html. R package version 0.2-4\/}.

\bibitem[\protect\citename{Zhang et~al., }2014]{zha14}
Zhang, X., Holt, J.~B., Lu, H., Wheaton, A.~G., Ford, E.~S., Greenlund, K.~J.,
  and Croft, J.~B. (2014).
\newblock \enquote{Multilevel regression and poststratification for small-area
  estimation of population health outcomes: a case study of chronic obstructive
  pulmonary disease prevalence using the behavioral risk factor surveillance
  system.}
\newblock {\em American Journal of Epidemiology\/}, 179, 8, 1025--1033.

\end{thebibliography}

\section*{Appendix: Full Conditional Distributions for PL-MB Model}

Let $\bm{\Omega} = \hbox{diag}(\omega_1,\ldots,\omega_n)$, and $\bm{\kappa}=\left(\tilde{w}_1*(y_1 - n_1/2), \ldots, \tilde{w}_n*(y_n - n_n/2) \right)'$. Note that $\bm{\kappa}/\bm{\omega}$ represents element-wise division.

%\begin{equation}
    $$\begin{aligned}
    \omega_i | \cdot & \sim \hbox{PG}(\tilde{w}_i*n_i, \; \bm{x}_i'\bm{\beta} + \bm{\psi}_i'\bm{\eta}), \; i=1,\ldots,n \\
            \bm{\eta} | \cdot & \propto  \prod_{i=1}^n \hbox{exp}\left(\kappa_i \bm{\phi_i'\eta} - \frac{1}{2}\omega_i(\bm{\phi_i'\eta})^2 - \omega_i (\bm{\phi_i'\eta})(\bm{x_i'\beta})\right) \\
            & \times \hbox{exp}\left(-\frac{1}{2\sigma^2_{\eta}}\bm{\eta'}\bm{\eta}\right) \\
            & \propto \hbox{exp}\left(-\frac{1}{2}(\bm{\kappa}/\bm{\omega} - \bm{X\beta} - \bm{\Phi}\bm{\eta} )' \bm{\Omega} (\bm{\kappa}/\bm{\omega}  - \bm{X\beta} - \bm{\Phi}\bm{\eta}) - \frac{1}{\sigma^2_{\eta}}\bm{\eta'}\bm{\eta} \right) \\
        \bm{\eta} | \cdot & \sim \hbox{N}_r\left(\bm{\mu} = (\bm{\Phi'\Omega \Phi} + \frac{1}{\sigma^2_{\eta}} \bm{I}_r)^{-1}\bm{\Phi'}\bm{\Omega}(\bm{\kappa}/\bm{\omega}-\bm{X \beta}), \;  \bm{\Sigma}=(\bm{\Phi'\Omega \Phi} + \frac{1}{\sigma^2_{\eta}} \bm{I}_r)^{-1}   \right) \\
    \end{aligned}$$

     $$\begin{aligned}
        \bm{\beta} | \cdot & \propto  \prod_{i=1}^n \hbox{exp}\left(\kappa_i \bm{x_i'\beta} - \frac{1}{2}\omega_i(\bm{x_i'\beta})^2 - \omega_i (\bm{x_i'\beta})(\bm{\phi_i'\eta})\right) \\
        & \times \hbox{exp}\left(-\frac{1}{2\sigma^2_{\beta}}\bm{\beta'}\bm{\beta}\right) \\
        & \propto \hbox{exp}\left(-\frac{1}{2}(\bm{\kappa}/\bm{\omega} - \bm{\Phi}\bm{\eta} - \bm{X\beta})' \bm{\Omega} (\bm{\kappa}/\bm{\omega} - \bm{\Phi}\bm{\eta} - \bm{X\beta}) - \frac{1}{\sigma^2_{\beta}}\bm{\beta'}\bm{\beta} \right) \\
        \bm{\beta} | \cdot & \sim \hbox{N}_p\left(\bm{\mu} = (\bm{X'\Omega X} + \frac{1}{\sigma^2_{\beta}} \bm{I}_p)^{-1}\bm{X'}\bm{\Omega}(\bm{\kappa}/\bm{\omega}-\bm{\Phi \eta}), \;  \bm{\Sigma}=(\bm{X'\Omega X} + \frac{1}{\sigma^2_{\beta}} \bm{I}_p)^{-1}   \right) \\
        \sigma^2_{\eta} | \cdot & \propto \left(\sigma^2_{\eta}\right)^{-\frac{r}{2}}\hbox{exp}\left(-\frac{1}{2\sigma^2_{\eta}}\bm{\eta'}\bm{\eta}\right)\\
        & \times \left(\sigma^2_{\eta}\right)^{-a - 1}\hbox{exp}\left(-\frac{1}{\sigma^2_{\eta}}b\right) \\
        & \propto \left(\sigma^2_{\eta}\right)^{-(a+\frac{r}{2}) - 1}\hbox{exp}\left(-\frac{1}{\sigma^2_{\eta}}(b + \frac{\bm{\eta}'\bm{\eta}}{2})\right) \\
        \sigma^2_{\eta} | \cdot & \sim \hbox{IG}(a + \frac{r}{2}, \; b + \frac{\bm{\eta}'\bm{\eta}}{2})
    \end{aligned}$$

%\end{equation}

\end{document}